\documentclass[10pt]{article}

\usepackage{amsmath}
\usepackage{amssymb}
\usepackage{graphics}
\usepackage{rotating}
\usepackage{cite}
\usepackage{color}
\usepackage{fancybox}

\def\0{\mbox{\tiny $0$}}
\def\1{\mbox{\tiny $1$}}
\def\2{\mbox{\tiny $2$}}
\def\3{\mbox{\tiny $3$}}
\def\4{\mbox{\tiny $4$}}
\def\5{\mbox{\tiny $5$}}
\def\6{\mbox{\tiny $6$}}
\def\7{\mbox{\tiny $7$}}
\def\8{\mbox{\tiny $8$}}
\def\9{\mbox{\tiny $9$}}
\def\I{\mbox{\tiny $I$}}
\def\II{\mbox{\tiny $II$}}
\def\III{\mbox{\tiny $III$}}
\def\L{\mbox{\tiny $L$}}
\def\D{\mbox{\tiny $D$}}
\def\T{\mbox{\tiny $T$}}

\def\max{\mbox{\tiny $max$}}
\def\min{\mbox{\tiny $min$}}
\def\tun{\mbox{\tiny $tun$}}

\title{\shadowbox{\large \bf  A STUDY OF TRANSIT TIMES IN DIRAC TUNNELLING}}

\author{
\small  Stefano De Leo\thanks{Department of Applied Mathematics,
State University of Campinas, Brazil [deleo@ime.unicamp.br] }
}

\date{\small
\fcolorbox{black}{yellow} {\color{red} $\bullet$ {\color{black}{
{\footnotesize  {\sc Journal of Physics A} {\bf 46}, 155306-11 (2013)}}}
{\color{red}{$\bullet$}} } }

\begin{document}
%

\maketitle

\vspace*{-.7cm}

\begin{abstract}
\noindent Based on the Dirac equation, the behavior  of relativistic electrons  which tunnel a
potential barrier of height $V_{\0}$ for  incoming energies
between $V_{\0}$ and $V_{\0}+m$ is studied by using the wave
packet formalism. The choice of this incoming energy zone
guarantees that only electrons participate to the tunneling
process. In the opaque limit, as shown in a previous analysis,  the
transit velocity is proportional to the barrier width and for
relativistic potential tends to a constant value greater than $c$.
In this paper, a {\em new} numerical study shows a very surprising result:
superluminal transmissions are also evident for thin barriers.
\end{abstract}











\section*{\normalsize I. INTRODUCTION}

In this paper, by using the Dirac equation, we examine the relationship between the peak of the
incoming wave packet (of energy $E$ and mass $m$) which encounters a potential barrier of
height $V_{\0}$ and width $L$ at $z=0$  and the peak of
transmitted wave packet which moves in the free potential region
after the barrier. The terminology ``Dirac tunneling'' refers to
the energy zone of the incoming particles,
$V_{\0}<E<V_{\0}+m$\cite{A1a,A1b,A1c}. This energy zone is
characterized by evanescent solutions in the region $0<z<L$, i.e.
$\exp[\pm \,\rho\,z]$ with $\rho=\sqrt{m^{\2}-(E-V_{\0})^{^{2}}}$.
The terminology Dirac tunneling is useful to distinguish these
evanescent solutions form the evanescent solutions which appear in
the Klein tunneling zone, $V_{\0}-m<E<V_{\0}$\cite{A1c}. The study
of the Klein tunneling zone exceeds the scope of this paper and it
will be appropriately discussed in a forthcoming article. It is
important to observe that the Klein tunneling zone has {\em not}
to be confused with the standard Klein  zone, $E<V_{\0}-m$, which
is characterized by oscillating antiparticle solutions, i.e.
$\exp[\pm \,i\,q\,z]$ with
$q=\sqrt{(E-V_{\0})^{^{2}}-m^{\2}}$ and the appearance of the Klein paradox\cite{A2,A3,A4,A5}.
 For the
convenience of the reader, we draw the energy Dirac/Klein zones in
the following picture

\begin{picture}(180,90)
 \put(48,0){$0$}\put(139,0){$L$}
\put(63,62){\mbox{\small \sc Oscillatory} $\,e^{-}$}
\put(63,48){\mbox{\small \sc Evanescent} $\,\,e^{-}$}
\put(63,36){\mbox{\small \sc Evanescent} $\,\,e^{+}$}
\put(63,22){\mbox{\small \sc Oscillatory} $\,e^{+}$}
 \put(13.7,71){\mbox{\small \sc I}}
 \put(13,64){\mbox{\small \sc n}}
 \put(13,57){\mbox{\small \sc c}}
 \put(13,50){\mbox{\small \sc o}}
 \put(13,43){\mbox{\small \sc m}}
 \put(14.5,36){\mbox{\small \sc i}}
 \put(13,29){\mbox{\small \sc n}}
 \put(13,22){\mbox{\small \sc g}}
\put(24.4,71){\mbox{\small \sc E}}
 \put(25,64){\mbox{\small \sc l}}
 \put(25,57){\mbox{\small \sc e}}
 \put(25,50){\mbox{\small \sc c}}
 \put(25,43){\mbox{\small \sc t}}
 \put(25,36){\mbox{\small \sc r}}
 \put(25,29){\mbox{\small \sc o}}
 \put(25,22){\mbox{\small \sc n}}
\put(25.5,15){\mbox{\small \sc s}}
 \put(166.8,71){\mbox{\small \sc O}}
 \put(168,64){\mbox{\small \sc u}}
 \put(168,57){\mbox{\small \sc t}}
 \put(168,50){\mbox{\small \sc g}}
 \put(168,43){\mbox{\small \sc o}}
 \put(169.5,36){\mbox{\small \sc i}}
 \put(168,29){\mbox{\small \sc n}}
 \put(168,22){\mbox{\small \sc g}}
\put(178,71){\mbox{\small \sc E}}
 \put(180,64){\mbox{\small \sc l}}
 \put(180,57){\mbox{\small \sc e}}
 \put(180,50){\mbox{\small \sc c}}
 \put(180,43){\mbox{\small \sc t}}
 \put(180,36){\mbox{\small \sc r}}
 \put(180,29){\mbox{\small \sc o}}
 \put(180,22){\mbox{\small \sc n}}
\put(180.5,15){\mbox{\small \sc s}} \put(5,-5){\mbox{\small \sc
region I}}\put(79,-5){\mbox{\small \sc region
II}}\put(160,-5){\mbox{\small \sc region III}}
 \thicklines
 \put(0,10){\line(1,0){50}}\put(140.5,10){\line(1,0){77}}
\put(50,10){\line(0,1){48}} \put(141,10){\line(0,1){48}}
\put(49.5,58){\line(1,0){92}}
 \put(0,57.){$....$}
  \put(0,44.5){$....$}
\put(33,57.){$......$} \put(33,44.5){$......$}
\put(140.5,57.){$.........$} \put(0,33){$....$}
\put(52,44.5){$.........................................$}
\put(33,33){$................................................$}
\put(187,57.){$...............................................................................$}
\put(187,33.){$...............................................................................$}
\put(187,44.5){$...............................................................................$}
 \put(340,64){\mbox{\small \sc Diffusion}} \put(270,64){$E>V_{\0}+m$}
 \put(340,48){\mbox{\small \sc
Dirac Tun}}
 \put(340,36){\mbox{\small \sc
Klein Tun }}
 \put(246,48){$V_{\0}<E<V_{\0}+m $}
  \put(225,36){$V_{\0}-m <E<V_{\0} $}
 \put(340,22){\mbox{\small \sc
Klein zone }} \put(270,22){$E<V_{\0}-m$}
\end{picture}
\vspace*{.5cm}

 \noindent The incoming electrons will be described by a wave packet
 with a  gaussian momentum distribution, $g(p)$  centered at
 $p_{\0}$,

 \[
g(p) = \left\{  \begin{array}{ll}
\exp\left[\,-\,(p-p_{\0})^{^{2}}d^{^{2}}\,/\,4\,\right]  &
\,\,\,\mbox{for}\,\,\,\,\,
\sqrt{V_{\0}^{^{2}}-m_e^{\2}}=p_{\min}\leq \,p\, \leq p_{\max}=\sqrt{V_{\0}(V_{\0}+2m_e)}\,\,,\\  \\
0 &\,\,\,\mbox{otherwise}\,\,,
\end{array} \right.
\]
where the minimum and maximum values of the momentum distribution
are chosen to guarantee Dirac tunneling.  The barrier filter
effect modifies the incoming momentum
distribution\cite{A6a,A6b,A6c,A6d}. The momentum mean value
$p_{\0}$ after transmission tends to an higher value, say
$p_{\L}$. In the opaque limit ($m_eL\gg1$) $p_{\L}\approx
p_{\max}$. For thin barriers $p_{\L} \approx p_{\0}$. The incoming
electrons move in the free region before the barrier with a
subluminal velocity given by $v_{\0}=p_{\0}/E_{\0}$. The outgoing
electrons move in the free region after the barrier with a
greater, but obviously still subluminal, velocity
$v_{\L}=p_{\L}/E_{\L}$. This implies that the maximum of the
spatial distribution of electrons, which tunnels a barrier of
width $L$, will reach the position $D>L$ at time
\begin{equation}
t_{\D,\L} = \frac{L}{v_{\tun}}\,+ \frac{D-L}{v_{\L}}\,\,.
\end{equation}
A natural question is to ask if and in which cases the average
velocity $v_{\D,\L}=D/\,t_{\L}$ is superluminal. Our paper was
intended as an attempt to give a satisfactory answer to this
intriguing question not only in the opaque limit, where the
transmission probability goes to zero, but also for thinner
barriers, where greater transmission probabilities are found.
Finally, the choice of the Dirac tunneling energy zone seems to be
the more appropriate in discussing superluminal velocities because
it avoids the possibility that the electrons which appear in the
free potential region after the barrier are products of pairs
production (a typical phenomenon of the Klein zone). The lower limit ($V_{\0}<m$) is not
covered in this work.

\section*{\normalsize II. THE DIRAC EQUATION IN PRESENCE OF A POTENTIAL BARRIER}

Since we shall need the solutions for a barrier potential, we
rewrite the free Dirac equation
\begin{equation}
i\,\gamma^{\mu}\partial_{\mu}\Psi(\boldsymbol{r},t)=m_e\,\Psi(\boldsymbol{r},t)\,\,,
\end{equation}
by including, via minimal coupling
$\partial_{\mu}\to\partial_{\mu}-i\,e\,A_{\mu}$\cite{A7a,A7b}, an
electrostatic potential
$A_{\mu}=\left[A(z)\,,\,\boldsymbol{0}\right]$, where
$A(z)=A_{\0}$ for $0<z<L$ and $0$ otherwise,
\begin{equation}
\left[\,i\,\gamma^{\mu}\partial_{\mu}+e\,\gamma_{\0}A(z)\,\right]
\Psi(\boldsymbol{r},t)=m_e\,\Psi(\boldsymbol{r},t)\,\,.
\end{equation}
Considering a one-dimensional motion, $p_x=p_y=0$ and $p_z=p$, and
stationary state solutions, $\psi(z)\exp[\,-\,i\,E\,t)]$, we
obtain
\begin{equation}
\gamma_{\0}\left[\,E-V(z)\,\right]\,\psi(z)+i\,\gamma_{\3}\,\psi'(z)=m_e\,\psi(z)\,\,,
\end{equation}
where $V(z)=-eA(z)$ and the prime indicates the derivative with
respect to $z$. In any constant potential region, for a given $E$,
only two solutions exist be they oscillatory ($|E-V_{\0}|>m$) or
evanescent ($|E-V_{\0}|<m$). Using the Pauli-Dirac set of gamma
matrices and observing that for one-dimensional phenomena spin
flip does not occur,  we find, for Dirac tunneling, the following
spinorial solutions\cite{A1c}
\begin{equation}
\begin{array}{lcl}
\mbox{\sc Region I} &\,\,\,:\,\,\, & u(p\,,\,E)\,\,e^{+\,i\,p\,z}
+R\,\,
u(-p\,,\,E)\,\,e^{-\,i\,p\,z}\,\,,\\
\mbox{\sc Region II} &\,\,\,:\,\,\, &
A\,\,u(i\,\rho\,,\,E-V_{\0})\,\,e^{-\,\rho\,z} +B\,\,
u(-\,i\,\rho\,,\,E-V_{\0})\,\,e^{+\,\rho\,z}\,\,,\\
\mbox{\sc Region III} &\,\,\,:\,\,\, &
T\,\,u(p\,,\,E)\,\,e^{+\,i\,p\,z} \,\,,
\end{array}
\end{equation}
where
\[ u(p\,,\,E) = \left[\,1\,,\,0\,,\,p\,/(E+m_e)\,,\,0\,\right]^{t}\,\,. \]
Because the Dirac equation is a first order equation in the
spatial derivatives, for a step-wise con\-ti\-nuous potential,
only the  continuity of the wave function has to be required.
Imposing the continuity conditions, $\psi_{\I}(0)=\psi_{\II}(0)$
and $\psi_{\II}(L)=\psi_{\III}(L)$, we obtain the following
transmission coefficient
\begin{equation}
\label{trac}
T(p,L)=e^{-\,i\,p\,L}\,\mbox{\Large $/$}\,\left[\,\cosh(\,\rho
L)\,-\,i\,\,\frac{p^{\2}-V_{\0}E}{p\,\rho}\, \sinh(\,\rho
L)\right]\,\,.
\end{equation}
The transmitted wave packet is then obtained by integrating over
all the possible stationary states modulated by the weighting
function $g(p)$,
\begin{eqnarray}
\label{traw} \Psi_{\T}(z,t) & = &
\int_{p_{\min}}^{p_{{max}}}\,\hspace*{-.4cm}\mbox{d}p\,\,g(p)\,T(p,L)\,u(p,E)\,\,
e^{i\,[\,p\,z-E\,t\,]} \nonumber \\ & = &
\int_{p_{\min}}^{p_{{max}}}\,\hspace*{-.4cm}\mbox{d}p\,\left[\,g_{\T}(p,L)\,,\,0\,,\,
f_{\T}(p,L)\,,\,0\,\right]^{t}\,\,e^{i\,\theta}\,
\,e^{i\,[\,p\,(z-L)-E\,t\,]}\,\,,
\end{eqnarray}
where
\[\tan\theta =
\frac{p^{\2}-V_{\0}E}{p\,\rho}\,\tanh(\rho L)\,\,\,,\,\,\,\,\,\,\,
g_{\T}(p,L)=g(p)\,\,|T(p,L)|\,\,\,\,\,\,\,\mbox{and}\,\,\,\,\,\,\,
f_{\T}(p,L)=\frac{p}{E+m_e}\,\,g_{\T}(p,L)\,\,.\]

\section*{\normalsize III. THE FILTER EFFECT FOR OPAQUE BARRIERS}

This section contains a brief summary of the discussion recently appeared in literature\cite{PRA83} on the filter effect for opaque barriers in the Dirac equation.  The action of the filter effect\cite{A6a} on the momentum
distribution of the transmitted wave, $g_{\T}(p,L)$ and
$f_{\T}(p,L)$, is shown in Fig.\,1. For increasing values of the
barrier width $L$, the transmitted momentum averaged over  the
distributions $g_{\T}(p,L)$ and $f_{\T}(p,L)$,
\begin{equation}
p_{\L}:=
\langle p \rangle_{\T} =
\int_{p_{\min}}^{p_{{max}}}\,\hspace*{-.4cm}\mbox{d}p\,p\,\left[\,g^{\2}_{\T}(p,L)+
f^{^{2}}_{\T}(p,L)\right] \,\mbox{\huge
/}\,\int_{p_{\min}}^{p_{{max}}}\,\hspace*{-.4cm}\mbox{d}p\,\left[\,g^{\2}_{\T}(p,L)+
f^{^{2}}_{\T}(p,L)\right]\,\,,
\end{equation}
  tends to
$p_{\max}$. This splitting of the average momentum is due to the
higher momenta selection caused by the exponential functional
dependence, $\exp[-\rho L]$, of the transmission coefficient
$T(p,L)$. The plots in Fig.\,1 refer to a potential barrier of
height $V_{\0}=m_e$, which implies $p_{\min}=0$ and
$p_{\max}=\sqrt{3}\,m_e$, and an incoming momentum
$p_{\0}=\sqrt{3}\,m_e/2$. The space localization for the incoming
wave packet is determined by $m_ed=10$. The ratio between the down
and up distribution, as a consequence of the higher momenta
filter, increases from $p_{\0}/(E_{\0}+m_e)\approx 37.3\%$ to
$p_{\max}/(E_{max}+m_e)\approx 57.7\%$.

Without going into the details of the mathematical apparatus of
asymptotic integral expansions and, in particular, of the validity
and possible generalization of the stationary phase method which
find a complete description in a number of books\cite{A8a,A8b} and
papers\cite{A6a,A6b,A6c,A6d}, we briefly discuss the case of {\em
opaque barriers}, i.e. $m_e L\gg 1$. In this limit, the cosh and
sinh functions with argument $\rho L$ can be approximated by
$\exp[-\rho L]/2$. This allows to rewrite  the transmitted
coefficient  as
\begin{equation}
|T(p,m_eL\gg 1)|\approx
2\,p\,\rho\,e^{-\,\rho\,L}\,/\,m_eV_{\0}\,\,.
\end{equation}
Now, the computation of the integral given in Eq.\,(\ref{traw}),
taking
 into account the previous approximation, can be done
 analytically. Starting from
\begin{equation}
||\Psi_{\T}(L,t;m_eL\gg 1)\,||  \approx \frac{2}{m_e
V_{\0}}\,\,g(p_{\max})\,\sqrt{\frac{2\,E_{\max}}{E_{\max}+m_e}}\,\,\,\left|\,
\int_{p_{\min}}^{p_{\max}}\,\hspace*{-.4cm}\mbox{d}p\,\,p\,\rho\,
\,\, e^{-\,\rho\,L}\,e^{i\,(\theta-E\,t)}\right|\,\,,
\end{equation}
changing the variable of integration from $p$ to $\rho$,
\[ p\,\mbox{d}p \approx -
\,\frac{E_{max}}{V_{\0}-E_{max}}\,\rho\,\mbox{d}\rho\,\,\,,\,\,\,\,\,\,\,
\rho(p_{\min})=m_e\,\,\,\,\,\mbox{and}\,\,\,\,\,\rho(p_{\max})=0\,\,,\]
and using the following series expansions for $\theta$ and $E$,
\[
\theta \approx  \theta_{\max} -\,a_{\1} \,\rho \,+\,
\mbox{O}[\,\rho^{\3}]\,\,\,\,\,\mbox{and}\,\,\,\,\, E  = E_{\max}
-\,a_{\2}\,\rho^{\2}\,+\, \mbox{O}[\,\rho^{\4}]\,\,,
\]
where $a_{\1}=\sqrt{V_{\0}(V_{\0}+2\,m_e)}\,/m_eV_{\0}$ and
$a_{\2}=1/2m_{e}$, we find
\begin{eqnarray}
 ||\Psi_{\T}(L,t;m_eL\gg 1)\,|| & \propto &\left|\,
\int_{0}^{m_e}\,\hspace*{-.4cm}\mbox{d}\rho\,\,\rho^{\2}\,e^{-\,\rho
L} \,\, \exp\left[\,i\,\left(\,a_{\2}t\,\rho^{\2} -
\,a_{\1}\,\rho\,\right)
\right]\right| \nonumber \\
&\approx & \left|\,
\int_{0}^{m_e}\,\hspace*{-.4cm}\mbox{d}\rho\,\,\rho^{\2}\,e^{-\,\rho
L} \,\, \left[\,1 +i\,\left(\,a_{\2}t\,\rho^{\2}
-a_{\1}\,\rho\,\right) - \left(\,a_{\2}t\,\rho^{\2}
-a_{\1}\,\rho\,\right)^{^{2}}/\,2   \right]\right|\,\,.
\end{eqnarray}
It is clear from this discussion that the problem of finding the
maximum of the transmitted wave packet reduces, in the opaque
barrier limit,  to find the maximum of the function
\begin{equation}
\mathcal{P}_{\T}(t)=\left|\,s(2)-\left[(a_{\2}t)^{\2}\,s(6)+a_{\1}^{\2}s(4)\right]/\,2
 +a_{\1}a_{\2}t\,s(5)+i\,\left[a_{\2}t\,s(4)-a_{\1}s(3)\right]\,\right|^{^{2}}\,\,,
 \end{equation}
where
\[
s(n)=\int_{\0}^{m_e}\,\hspace*{-.2cm}\mbox{d}\rho\,
\rho^{n}\,e^{-\,\rho L} \approx\,\, n!\,/\,L^{^{n+1}}\,\,.
\]
The condition $\partial_t||\Psi_{\T}(L,t;m_eL\gg 1)||^{^{2}}=0$ is
thus satisfied requiring $\mathcal{P}_{\T}'(t)=0$. Observing that
\[
\mathcal{P}_{\T}(t)\approx
 s^{\2}(2) + a_{\1}^{\2}\, [\,s^{\2}(3)-s(2)s(4)\,]+
2\,a_{\1}a_{\2}t\,[\,s(2)s(5)-s(3)s(4)\,]+ (a_{\2}t)^{\2}
[\,s^{\2}(4)-s(2)s(6)\,]\,\,,
\]
we find for the tunneling time, $\tau_{\tun}$, the following
equation
\begin{equation}
a_{\1}[\underbrace{s(2)s(5)-s(3)s(4)}_{4\,\cdot
\,4!\,/\,L^{^{9}}}]+\,
a_{\2}\tau_{\tun}\,[\underbrace{s^{\2}(4)-s(2)s(6)}_{-\,36\,\cdot\,4!\,/\,L^{^{10}}}]=0\,\,.
\end{equation}
From this equation, we obtain a  closed formula for the tunneling
velocity,
\begin{equation}
\label{vtun} v_{\tun}=\frac{L}{\tau_{\tun}}=
\frac{9\,a_{\2}}{a_{\1}}=\,
\frac{9}{2}\,\sqrt{\frac{V_{\0}}{V_{\0}+2\,m_e}}\,\,.
\end{equation}
In order to understand the importance of Eq.\,(\ref{vtun}), let us
present two limit cases. For non relativistic potentials,
\[v_{\tun}[V_{\0}\ll m_e]\approx \frac{9}{2}\,\sqrt{\frac{V_{\0}}{2\,m_e}}\ll
1\,\,.
\]
This means that the use the Schr\"odinger equation leads to
non-relativistic tunneling velocities. A surprising result is seen
in the chiral limit,
\[v_{\tun}[V_{\0}\gg m_e]\approx \frac{9}{2}\,\,,
\]
where superluminal tunneling velocities appear\cite{PRA83}.

\section*{\normalsize IV. NUMERICAL ANALYSIS AND TRANSIT TIMES}

 Up to this point,
we have limited ourselves to a strictly analytic approach to the
problem. To test our analysis, we have performed a numerical study by using
{\em Mathematica 8} [Wolfram]
whose results are shown in Fig.\,2 and Fig.\,3. The numerical
analysis shows the phenomenon of multiple peaks. The spatial
distribution of these peaks is plotted in Fig.\,2 for a barrier of
height $V_{\0}=m_{e}$ and incoming momentum distribution centered
to $p_{\0}=\sqrt{3}\,m_{e}/\,2$. In Table 1, we give the values of
the transmitted probability density  and the corresponding
appearance time for the first peaks around the main maximum. For
thinner barrier the secondary peaks are very small when compared
to the main peak. The phenomenon of multiple peaks disappears for
increasing values of the barrier width. In Fig.\,3, we show the
barrier width dependence of the tunneling time, $\tau_{\tun}$,
referring to the main peak and the corresponding tunneling
velocity, $v_{\tun}$. The analytic value of the tunneling
velocity, calculated by using Eq.\,(\ref{vtun}) for $V_{\0}=m_e$,
$9/2\sqrt{3}$, is in good agreement with the numerical data. The
important point to be noted here is that the numerical data
clearly show superluminal tunneling velocities also for {\em
thinner} barriers, see Fig.\,3 in the initial range of the barrier
width. The conclusion of our numerical study is unexpected and
quite surprising. For thinner barrier superluminal tunneling
velocities appear and the superluminal effects are {\em greater}
than those observed in the opaque barrier limit. Due to this
surprising result, we have investigated in details the behavior of
an incoming wave packet of momentum $p_{\0}=\sqrt{3}\,m_e/\,2$ and
localization $d=10/m_e$ which starting from the point $z=0$ at
time $t=0$ reaches the point $D=40/m_e$ travelling through free
space (Fig.\,4a), tunneling a barrier of width $L=10/m_e$
(Fig.\,4b), a barrier of width $L=20/m_e$ (Fig.\,4c), and a
barrier of width $L=30/m_e$ (Fig.\,4d). The transmitted peak is
found at $D=40/m_e$ at the following times, $\{t_{\D,\L}\}$,
\[
\left\{\,m_e\,t_{\4\0,\0}\,,\,m_e\,t_{\4\0,\1\0}\,,\,m_e\,t_{\4\0,\2\0}\,,
\,m_e\,t_{\4\0,\3\0}\,\right\}
=\left\{\,61.35\,,\,46.03\,,\,29.95\,,\,15.70\,\right\}\,\,.
\]
These numerical times  imply the following transit velocities,
$\{v_{\D,\L}\}$,
\[
\left\{\,v_{\4\0,\0}\,,\,v_{\4\0,\1\0}\,,\,v_{\4\0,\2\0}\,,\,v_{\4\0,\3\0}\,\right\}
=\left\{\,0.65\,,\,0.87\,,\,1.34\,,\,2.55\,\right\}\,\,.
\]
The condition for superluminality in $D$ is
\[
v{\D,\L}=\frac{D\,v_{\tun}\,v_{\L}}{L\,v_{\L}+(D-L)\,v_{\tun}}>1\,\,\,\,\,\,\,\Rightarrow
\,\,\,\,\,\,\, D <
\frac{v_{\tun}-v_{\L}}{v_{\tun}(\,1-v_{\L})}\,\,L\,\,.
\]

\section*{\normalsize V. CONCLUSIONS}

The discussion of the time spent by a particle to tunnel through a barrier potential has been approached from many different points of view and has been the issue of an intriguing debate in the last decades\cite{A6c,A6d}. Since the energy is bounded from below, the time operator\cite{TO}
is not self-adjoint, there is no general agreement about a satisfactory definition   of tunneling times and a universal intrinsic tunneling time  valid for all experiments probably does not exist.

The {\em phase time}  is the time between the instant in which the peak of the incoming wave packet reaches the barrier and the instant in which the peak of the transmitted wave packet  appears in the free region after the barrier. By using the stationary phase method, Hartman\cite{A6a} observed that the tunneling thorugh opaque barriers is independent of the barrier length.  This surprising result is due to the fact that the stationary phase method was applied without considering the  {\em filter effect}.  In the  study presented in  ref.\cite{PRA83} and briefly summarized in section III,  the integral of the transmitted wave function for Dirac particles  was calculated taking into account the filter effect. The {\em new} formula for the tunneling velocity in presence of opaque barriers,
\[ v_{_{tun}}^{^{op.bar.}}=\,
\frac{9}{2}\,\sqrt{\frac{V_{\0}}{V_{\0}+2\,m_e c^{^{2}}}}\,\,c\,\,,
\]
clearly shows that  for non-relativistic potentials, $V_{\0}\ll m_ec^{\,\2}$,
the tunneling velocities are
subluminal. There is {\em not}\, Hartman effect for
the Schr\"odinger equation. Nevertheless, ``superluminality" for the phase time is seen for
relativistic potentials. For example, for $V_{\0}=m_ec^{\,\2}$ the
tunneling velocity, given by $9\,c\,/\,2\sqrt{3}$\,, is superluminal and
this  is confirmed by  numerical calculations, see Fig.\,2. The
numerical data also present a still more surprising result. In
Fig\,3, it is clear that superluminal tunneling velocities  appear
not only in the opaque limit (very wide barriers) but also for
{\em thinner} barriers.  This stimulated a
numerical analysis in which the transmission across free space, see Fig.\,4.a, and through
potential barriers of different widths, see
Fig.\,4.b-c, is compared. Superluminal transmissions appear.
This apparent paradox clearly shows that the phase time is {\em not} the correct tunneling time definition. Why? The peak of the transmitted wave packet could be {\em not} related to the peak of
the incoming one. When a wave packet with a given momentum distribution strikes a barrier, the transmitted wave packet will exhibit a distribution displaced to higher momenta. Consequently, the transmitted packet moves faster that the incident packet\cite{PRA83}. Observing  that
the transmitted coefficient only depends on the barrier width and not on its position, see Eq.(\ref{trac}),  the transit time does not change if we displace the barrier from ($0$,\,$L$) to ($a$,\,$a+L$).  This means, for example,  that  the time in which the peak of the transmitted wave packet reaches the point $z=40\,\hbar/m_e\,c$ in Fig.\,4 is the same for  barriers  located at $z=0$ or at $z>0$. Due to the fact that for $z>0$ the free region after the barrier is reduced, the outgoing peak cannot be related to the incoming one. Once again the {\em filter effect} explains the apparent paradox. Only higher momenta in the incoming packet contribute to tunneling.

In view of the last comment, it  could be interesting to review the results reported in this paper
in terms of other definitions of tunneling times. The {\em dwell time}\cite{LA} is defined as the ratio of the number of particles within the barrier to the incident flux. The average of the time spent by the particle inside the barrier obviously does not distinguish whether, at the end of the process, the particles are reflected or transmitted. Due to the fact that the effects shown in this paper imply a very small transmission, the average dwell time should be related, in this case, to the well-known  delay time of  quantum mechanics\cite{A8a} (obviously using the Dirac formalism it should contain the appropriated relativistic corrections). A more complicated analysis is required if we introduce a time-modulated barrier\cite{But}, $V(t)=V_{\0}+\widetilde{V}_{\0}\,\cos \omega t$. In non-relativistic quantum mechanics, the crossover between small and large frequencies yields to the  {\em traversal time}. The study of emitted and absorbed modulation quanta should be investigated in terms  of  relativistic transmitted particles. In this spirit, it could be also interesting to consider relativistic corrections to Larmor precession when $x$-polarized particles  which move in the $z$-direction encounter  in the barrier a small magnetic field pointing in the $y$-direction. The precession angle  determines the time spent by the particle in the barrier\cite{But}. Nevertheless, in recent papers, it was observed, for relativistic particles,  the phenomenon of spin\cite{spin} and helicity\cite{hel} flip for planar motion. Thus, it could be very interesting to investigate, for Dirac planar  motion,  possible consequences in the spin and helicity flip related to transversal times. We conclude this discussion with our last suggestions on possible future studies in which the numerical analysis  done in this paper could find possible applications. In recent  papers\cite{PT1,PT2}, the tunneling time problem for non-relativistic rectangular potential  barriers
was analyzed in terms of the {\em tempus} operator\cite{TO1,TO2} , $-i\,\hbar\,\partial_{_{E}}$.
The tunneling time is expressed by an energy average of the phase time and there is no fundamental problem with the Hartman effect\cite{PT1}. This conclusion is similar to our analysis for non-relativistic potentials.  In\cite{PT2}, it was shown that when the energy of the incoming particle is near to a resonance pole of the tunneling barrier, an intrinsic tunneling time does exist, but when the energy is near the branch point there is no intrinsic tunneling time. It could be interesting to
calculate the expectation value of the time operator, the resonance poles and the branch points for relativistic particles. Observing that the local value of the {\em tempus} operator gives a complex time for a particle to traverse a barrier, it should be also interesting to study the connection
between the results obtained in the presence time formalism and the results obtained by introducing
complex potential\cite{CP1,CP2}. In particular with respect to relativistic decoherence.

We hope that the  results presented in this
paper could stimulate and motivate studies  in tunneling phenomena by using the Dirac equation.
In this spirit, this work can be seen as an initial step in view of  new and more detailed discussions on the fascinating topic of relativistic tunneling.

\section*{\small \rm ACKNOWLEDGEMENTS}

The author  thanks the Department of Physics, University of
Salento (Lecce, Italy), for the hospitality and the FAPESP
(Brazil) for financial support by the Grant No. 10/02213-3. The author also thanks
the anonymous referees and the board member for their comments and suggestions.

\newpage

\begin{table}
\begin{center}
\shadowbox{
\begin{tabular}{|r|c||c||c|}
\hline     $m_e L$ & \multicolumn{3}{|c|}{{\sc Max}: $\left\{\,
m_e \,t\,,\,|\Psi_{\T}(L,t)|^{^{2}}\,\right\}$}  \\
 \hline \hline
 $10$ &
 $\left\{\,-65.97\,,\,1.73\times 10^{^{-20}}\,\right\}$ &
$\left\{\,\,\,2.05\,,\,2.29\times 10^{^{-8}}\,\,\,\right\}$  &
$\left\{\,71.14\,,\,2.69\times 10^{^{-20}}\,\right\}$
\\
\hline $15$ &
 $\left\{\,-47.96\,,\,3.24\times 10^{^{-21}}\,\right\}$ &
$\left\{\,\,2.06\,,\,2.49\times 10^{^{-12}}\,\,\right\}$  &
$\left\{\,53.06\,,\,4.39\times 10^{^{-21}}\,\right\}$
\\
\hline $20$ &
 $\left\{\,-34.50\,,\,1.05\times 10^{^{-22}}\,\right\}$ &
$\left\{\,\,2.15\,,\,3.17\times 10^{^{-16}}\,\,\right\}$  &
$\left\{\,39.76\,,\,1.30\times 10^{^{-22}}\,\right\}$
\\
\hline $25$ &
 $\left\{\,-18.15\,,\,5.83\times 10^{^{-22}}\,\right\}$ &
$\left\{\,\,2.63\,,\,6.29\times 10^{^{-20}}\,\,\right\}$  &
$\left\{\,24.02\,,\,5.55\times 10^{^{-22}}\,\right\}$
\\
\hline $30$ &
 $\left\{\,\,-8.64\,,\,1.12\times 10^{^{-22}}\,\,\right\}$ &
$\left\{\,\,3.22\,,\,2.10\times 10^{^{-22}}\,\,\right\}$  &
$\left\{\,14.74\,,\,1.20\times 10^{^{-22}}\,\right\}$
\\
\hline $40$ & $\star$ & $\left\{\,\,9.22\,,\,1.29\times
10^{^{-23}}\,\,\right\}$ & $\star$
\\
 \hline $50$ & $\star$ & $\left\{\,15.66\,,\,2.83\times 10^{^{-24}}\,\right\}$ &$\star$
\\
\hline $75$ &$\star$ & $\left\{\,26.44\,,\,2.14\times 10^{^{-25}}\,\right\}$ &
$\star$
\\
\hline $100$ & $\star$& $\left\{\,36.35\,,\,3.63\times 10^{^{-26}}\,\right\}$
&$\star$
\\
\hline
\end{tabular}
} \caption{Transmission probability densities calculated at $z=L$
and corresponding times in which the transmitted peaks appear in
$z=L$. For thin barriers, the secondary peaks are very small with
respect to the central peak. The phenomenon of multiple peaks
disappears for increasing values of the barrier width. The
numerical data of this Table was prepared by using an incoming
momentum distribution centered at $p_{\0}=\sqrt{3}\,m_e/2$, a
localization determined by $d=10/m_e$, and a barrier potential of
height $V_{\0}=m_e$.}
\end{center}
\end{table}

\newpage

\begin{figure}[hbp]
\hspace*{-2.5cm}
\includegraphics[width=19cm, height=22cm, angle=0]{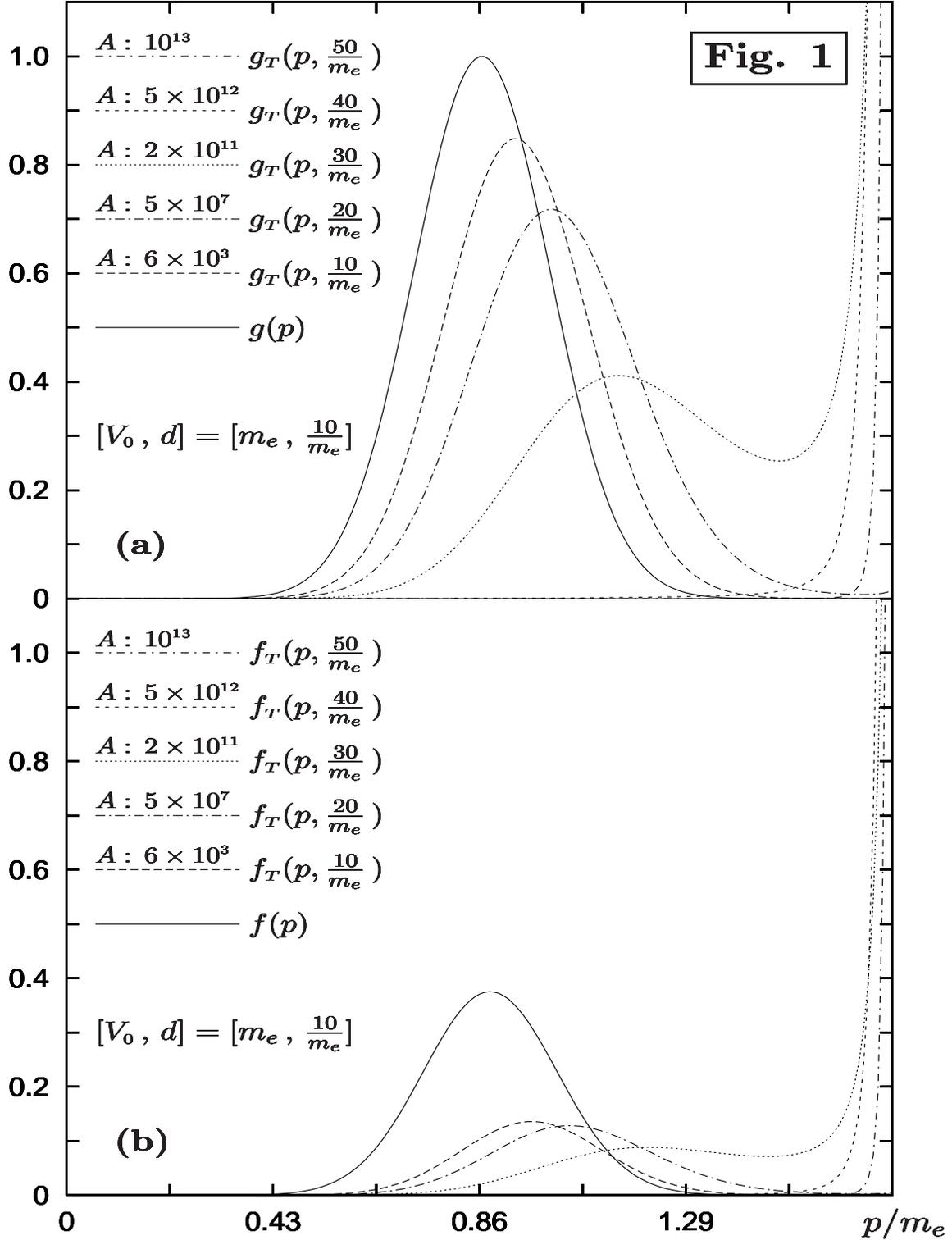}
\vspace*{-1.8cm}
 \caption{Momentum distributions for the up, $g_{\T}(p,L)$, and
down, $f_{\T}(p,L)$, component of the transmitted Dirac spinors.
The incoming momentum distribution $g(p)$ is centered  at
$p_{\0}=\sqrt{3}\,m_e/2$. Its space localization is determined by
$d=10/m_e$. By increasing the value of the barrier width $L$, the
filter effect modifies the initial momentum distributions leading,
for $m_eL\gg 1$, to smaller transmission probabilities, to greater
momentum mean values, $p_{\L}\to p_{\max}$, and to localizations
$\gg d$.}
\end{figure}

\newpage

\begin{figure}[hbp]
\hspace*{-2.5cm}
\includegraphics[width=19cm, height=22cm, angle=0]{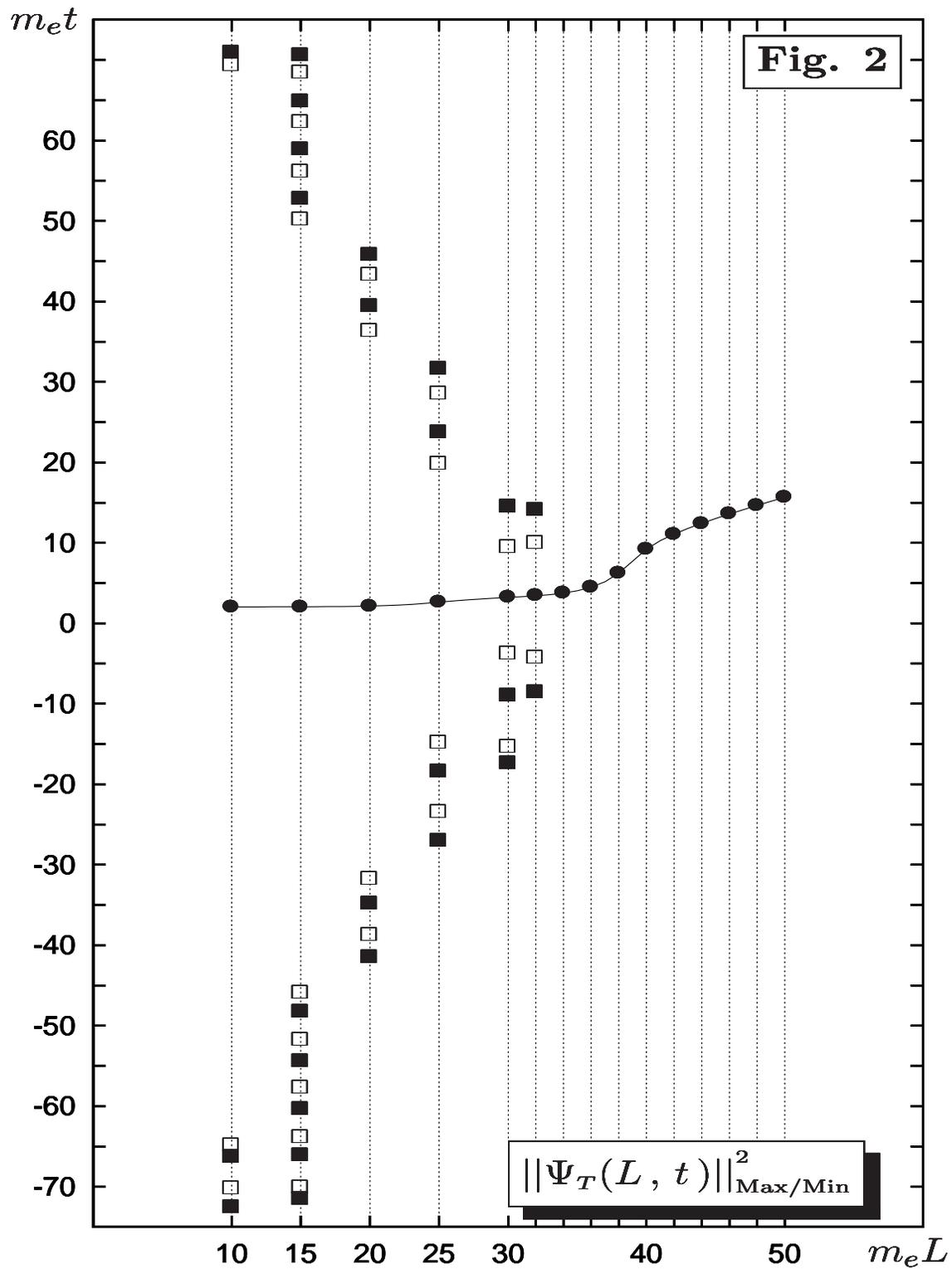}
\vspace*{-1.8cm}
 \caption{Maximum and minimum values of the transmitted probability density at $z=L$
in correspondence to the time in which it appears in the free
region after the barrier. The black circles indicates the central
peak. The secondary peaks are identified by the black squares.
Finally, the white squares represent the times in which the minima
appear in $z=L$. The numerical data refer to incoming electrons,
with a momentum distribution centered at $p_{\0}=\sqrt{3}\,m_e/2$
 and localization determined by $d=10/m_e$,
 which tunnel a potential barrier of height $V_{\0}=m_e$ and width $L$.}
\end{figure}

\newpage

\begin{figure}[hbp]
\hspace*{-2.5cm}
\includegraphics[width=19cm, height=22cm, angle=0]{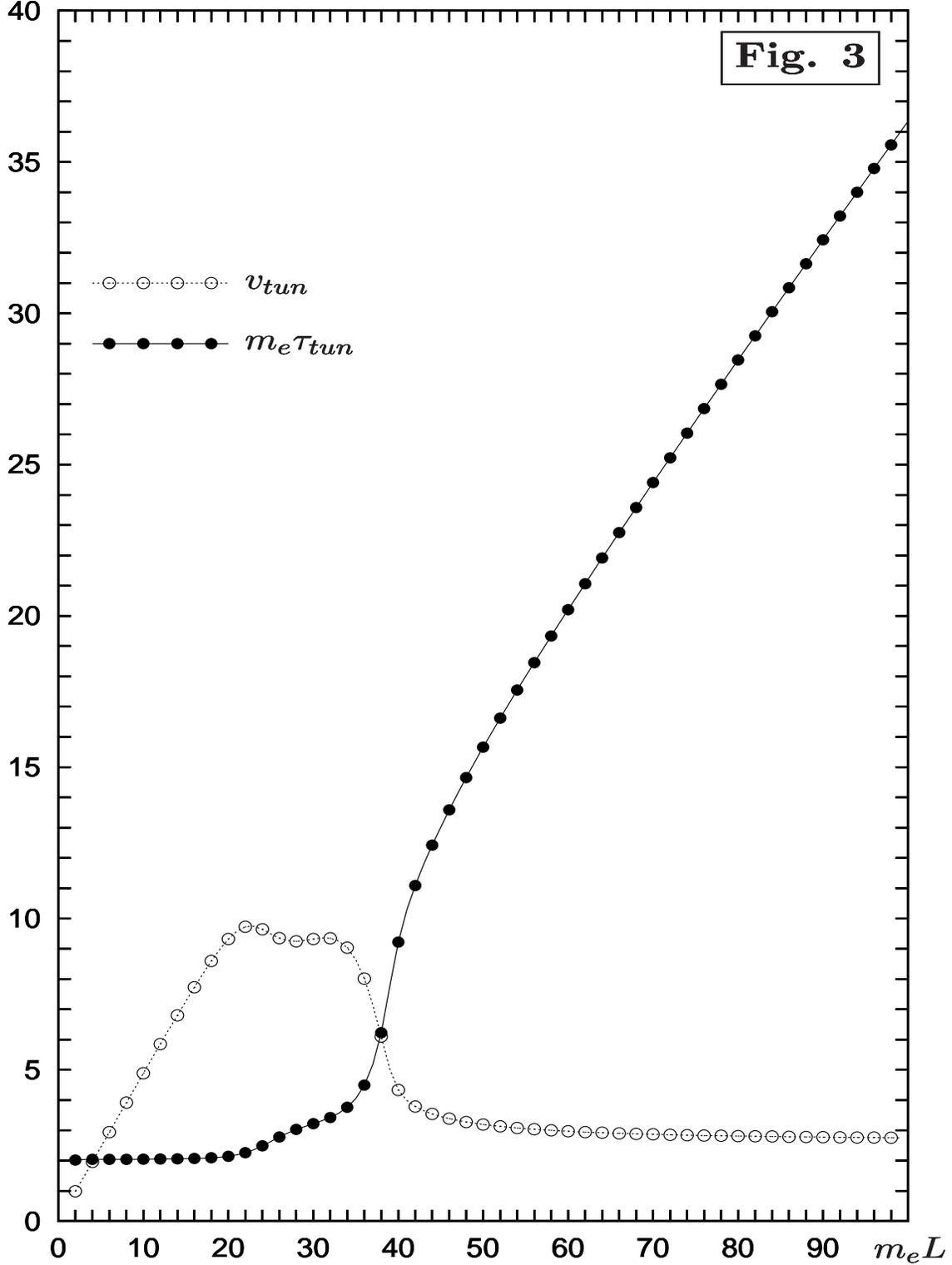}
\vspace*{-1.8cm}
 \caption{Tunneling times (black circles and continuous line) and tunneling velocity
 (withe circles and dotted line) for incoming electrons of momentum
 $p_{\0}=\sqrt{3}\,m_{e}/2$ and localization $d=10/m_e $ which tunnel
 a potential barrier of height $V_{\0}=m_e$ and width $L$. The numerical data
 confirm the analytic formula obtained in the opaque limit for the tunneling velocity, see
 Eq.\,(\ref{vtun}). An
 unexpected and surprising result is seen for thin barriers ($\approx
 2\,d$) where  the superluminal tunneling velocity reaches its maximum value ($\approx
 10$).}
\end{figure}

\newpage

\begin{figure}[hbp]
\hspace*{-2.5cm}
\includegraphics[width=19cm, height=22cm, angle=0]{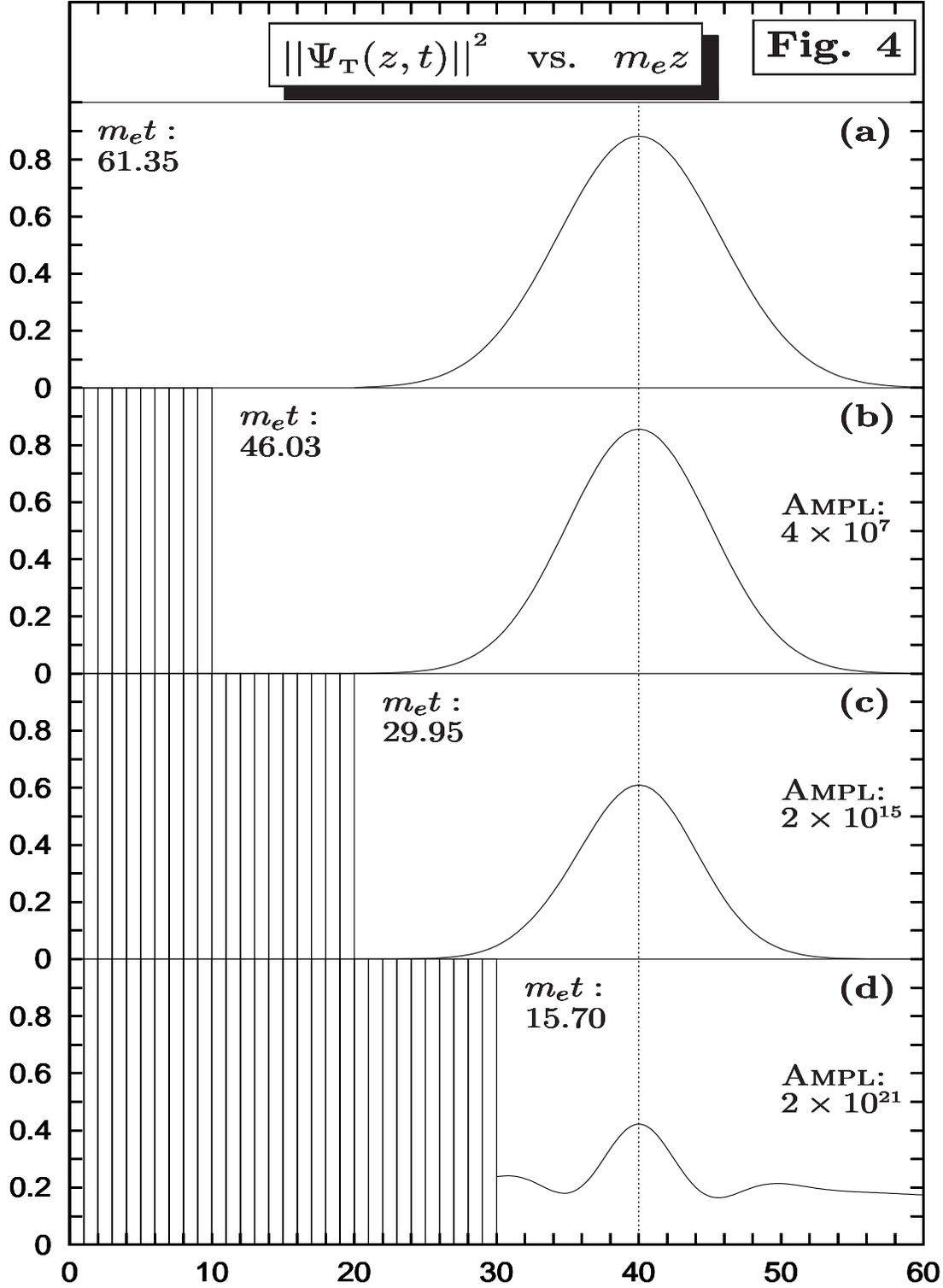}
\vspace*{-1.8cm}
 \caption{Probability density for incoming electrons, with a momentum distribution
 centered at
 $p_{\0}=\sqrt{3}\,m_e/2$ and localization $d=10/m_e$, which move in free space
 (a),   tunnel a potential barrier of width $L$ equal to $10/m_e$ (b), $20/m_e$ (c),
  and $30/m_e$ (d). In the free space the electrons move with a mean velocity
  $v_{\4\0,\0}\approx 0.65$. Barrier tunneling increases
  the velocity of propagation leading to superluminal transmission, $v_{\4\0,\2\0}\approx 1.34$
  and $v_{\4\0,\3\0}\approx 2.55$.}
\end{figure}

\end{document}